\documentclass[12pt,twoside]{article}
\usepackage{amsmath,fleqn,espcrc1}

\usepackage[dvips]{graphics,graphicx}

\title{Two-particle correlation from a relativistic fluid with a first order
phase transition}

\author{Kenji Morita\address{Department of Physics, Waseda University,\\
		Tokyo 169-8555, Japan}
        , Shin Muroya\address{Tokuyama Women's College,\\
		Tokuyama, Yamaguchi 745-8511, Japan}
		, Hiroki Nakamura$^{\rm a}$
		and
        Chiho Nonaka\address{Department of Physics, Hiroshima University,\\
		Higashi-hiroshima, Hiroshima 739-8526, Japan}}
       
\begin{document}

\maketitle

\setlength{\unitlength}{1cm}
\begin{picture}(0,0)(0,0)
 \put(13.93,8.7){WU-HEP-00-6}
 \put(14.63,8.2){TWC-00-5}
\end{picture}

\begin{abstract}
 We numerically calculate two-particle correlation functions of CERN-SPS 158
 A GeV Pb+Pb central collisions based on a (3+1)-dimensional relativistic
 hydrodynamics with a first order phase transition. We analyze the pair
 momentum dependence of pion source sizes extracted from the
 Yano-Koonin-Podgoretski\u{\i} (YKP) parametrization which is expected to
 give the source sizes directly. We find that, even in the case of the first
 order phase transition, the collective expansion and surface dominant
 freeze-out of the fluid naturally lead to the opaque source for which the 
 interpretation of the temporal source parameter as the emission duration
 breaks down. 
\end{abstract}

\section{INTRODUCTION}

In the physics of relativistic heavy ion collisions aiming at studying the
QCD phase transition of hot and dense matter, two-particle correlation is
widely used to investigate the characteristics of the space-time evolution in
the collision processes \cite{review}. It is well known as Hanbury-Brown
Twiss effect that the two-particle intensity correlation function contains
the information on the ``size'' of the source of emitted particles due to a
quantum symmetry. In recent experiments, the sizes are estimated from the
correlation function through Gaussian fitting and the physical meaning of
the sizes depends on the fitting function. For an azimuthally symmetric
source, the Yano-Koonin-Podgoretski\u{\i}(YKP) parametrization \cite{ykp} is
used as one of the fitting functions. In this parametrization, three of
fitting parameters can be regarded as source sizes directly. This
advantage gives us a chance to observe a signature of quark-gluon matter
because large emission duration is expected in the case of a first order
phase transition \cite{pratt}. In the case of the relativistic heavy ion
collisions, the reaction is highly dynamic; the particle source cannot be
considered to be static one. So we analyze the size parameters for pions
quantitatively based on a hydrodynamic model which takes into account the
longitudinal and transverse expansion and a first order phase transition
\cite{paper}.

\section{FORMALISM}

 From a hydrodynamic point of view, we consider the pion source as a
 thermalized one and a completely chaotic source. Hence, the
 two-particle correlation function is given as 
 \begin{equation}
  C(\mathbf{k_1,k_2})
   =1+\frac{|I(\mathbf{k_1,k_2})|^2}{(dN/d^3\mathbf{k_1})(dN/d^3\mathbf{k_2})},
   \label{c2}
 \end{equation}
 where $\mathbf{k_1}$ and $\mathbf{k_2}$ are momenta of a emitted pion pair
 and $I(\mathbf{k_1,k_2})$ is the interference term. 

 As usual, we introduce the relative momentum $q^\mu=k_1^\mu-k_2^\mu$ and
 the average momentum $K^\mu=(k_1^\mu+k_2^\mu)/2$ which can be put as
 $K^\mu=(K^0,K_T,0,K_L)$ by virtue of the cylindrical symmetry. Due to the
 on-shell condition, $q_\mu K^\mu=0$, only three components of the relative
 momenta are independent. In the YKP parametrization,
 $q_\bot=\sqrt{q_x^2+q_y^2}$, $q_\|=q_z$, and $q^0$ are used as the
 independent components. Then, the fitting function is given by
\begin{multline}
\!\!\!\! C(q^\mu,K^\mu)= 1+\lambda\exp 
  \left\{ 
   -R_\bot^2(K^\mu)q_\bot^2-R_\|^2(K^\mu)[q_\|^2-(q^0)^2]
\right.\\
 \qquad \qquad \qquad \qquad \left. -\left[
	 R_0^2(K^\mu)+R_\|^2(K^\mu)
	 \right]
 \left[
  q_\mu u^\mu (K^\mu)
  \right]^2 
  \right\} ,
\label{ykp}
\end{multline}
where $u^\mu$ is a four-velocity which has only longitudinal component.
The size parameters to be determined by a fit are $R_i$s $(i=\bot, \|,
0)$. We may choose a special reference frame where the velocity parameter $v$
vanishes (YKP frame), because the three size parameters are invariant under
longitudinal boosts. Introducing the source function,
\begin{equation}
 S(x^\mu, K^\mu)=\int_\Sigma \frac{U_\mu(x^\prime)d\sigma^\mu(x^\prime)}
  {(2\pi)^3}
  \frac{U_\nu(x^\prime)k^\nu}{\exp(U_\rho(x^\prime)k^\rho/T)-1}
  \delta^4(x-x^\prime), \label{S} 
\end{equation}
and the weighted average,
\begin{equation}
 \langle A(x^\mu) \rangle =\frac{\int d^4x A(x^\mu) S(x^\mu,K^\mu)}
  {\int d^4 x S(x^\mu,K^\mu)},
\end{equation}
the size parameters in this frame (YKP frame) are expressed as 
\begin{eqnarray}
 R_\bot^2(K^\mu)&=& (\Delta y)^2,\label{rt}\\
 R_0^2(K^\mu)&=& (\Delta t)^2
  -\frac{2}{\beta_\bot}\langle \tilde{x}\tilde{t} \rangle
  +\frac{1}{\beta_\bot^2}[(\Delta x)^2-(\Delta y)^2],\label{r0}\\
 R_\|^2(K^\mu)&=&(\Delta z)^2
  -\frac{2\beta_\|}{\beta_\bot}\langle \tilde{x}\tilde{z} \rangle
  +\frac{\beta_\|^2}{\beta_\bot^2}[(\Delta x)^2-(\Delta y)^2],\label{rl}
\end{eqnarray}
where $\tilde{x}=x-\langle x \rangle$, $\Delta x=\sqrt{\langle x^2 
\rangle-\langle x \rangle^2}$, $\beta_\bot=K_T/K^0$ and $\beta_\|=K_L/K^0$.

In order to regard the above $R_i$s as naive source sizes directly, the
second term and the third term in the right hand side of Eqs.(\ref{r0}) and
(\ref{rl}) must be small compared to the first term of each equation. Though
it has been shown that this condition holds within a class of thermal model
\cite{ykp}, it is not trivial in other cases. The case of a hydrodynamic
model as a more realistic one is the point of the present paper. In the
following, we calculate three quantities: HBT radii extracted from the fit to
Eq.(\ref{c2}), the space-time extensions, Eqs.(\ref{rt})-(\ref{rl}), and the
source sizes, the first terms in Eqs.(\ref{r0}) and (\ref{rl}).

\section{DISCUSSION}
 \begin{minipage}[c]{13cm}
  \baselineskip=12pt
  \begin{center}
   \includegraphics[scale=0.34]{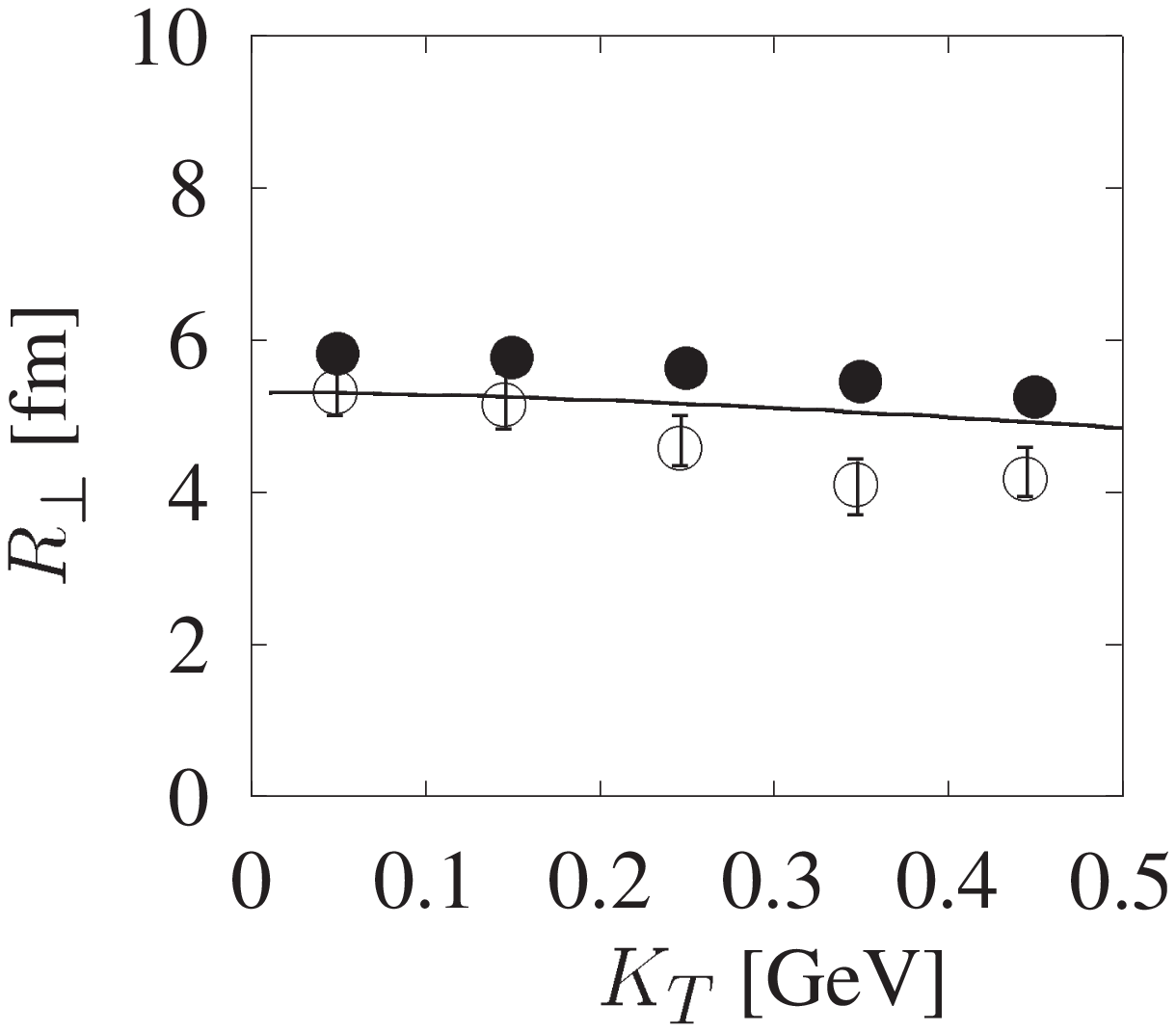}\,
   \includegraphics[scale=0.34]{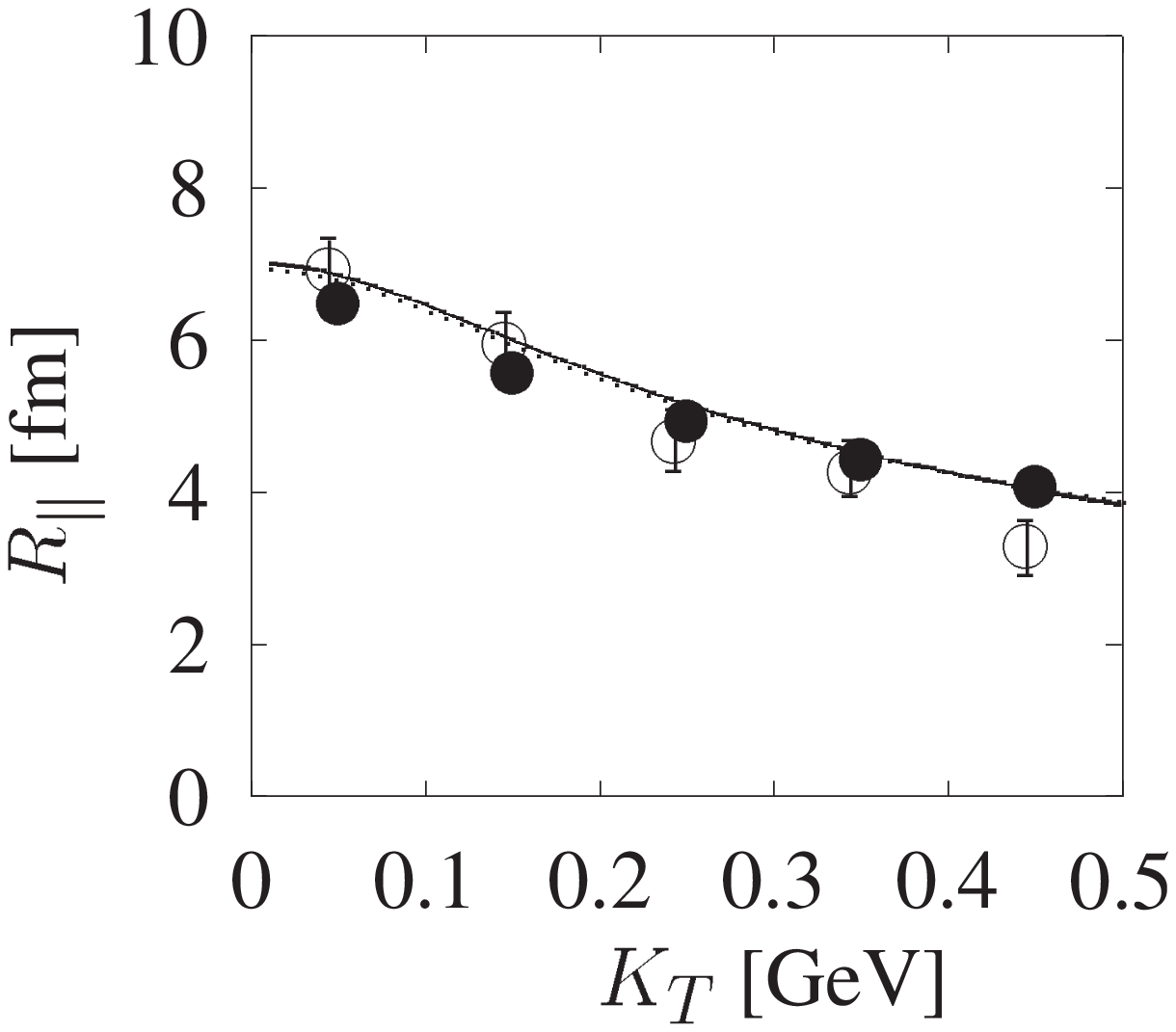}\,
   \includegraphics[scale=0.34]{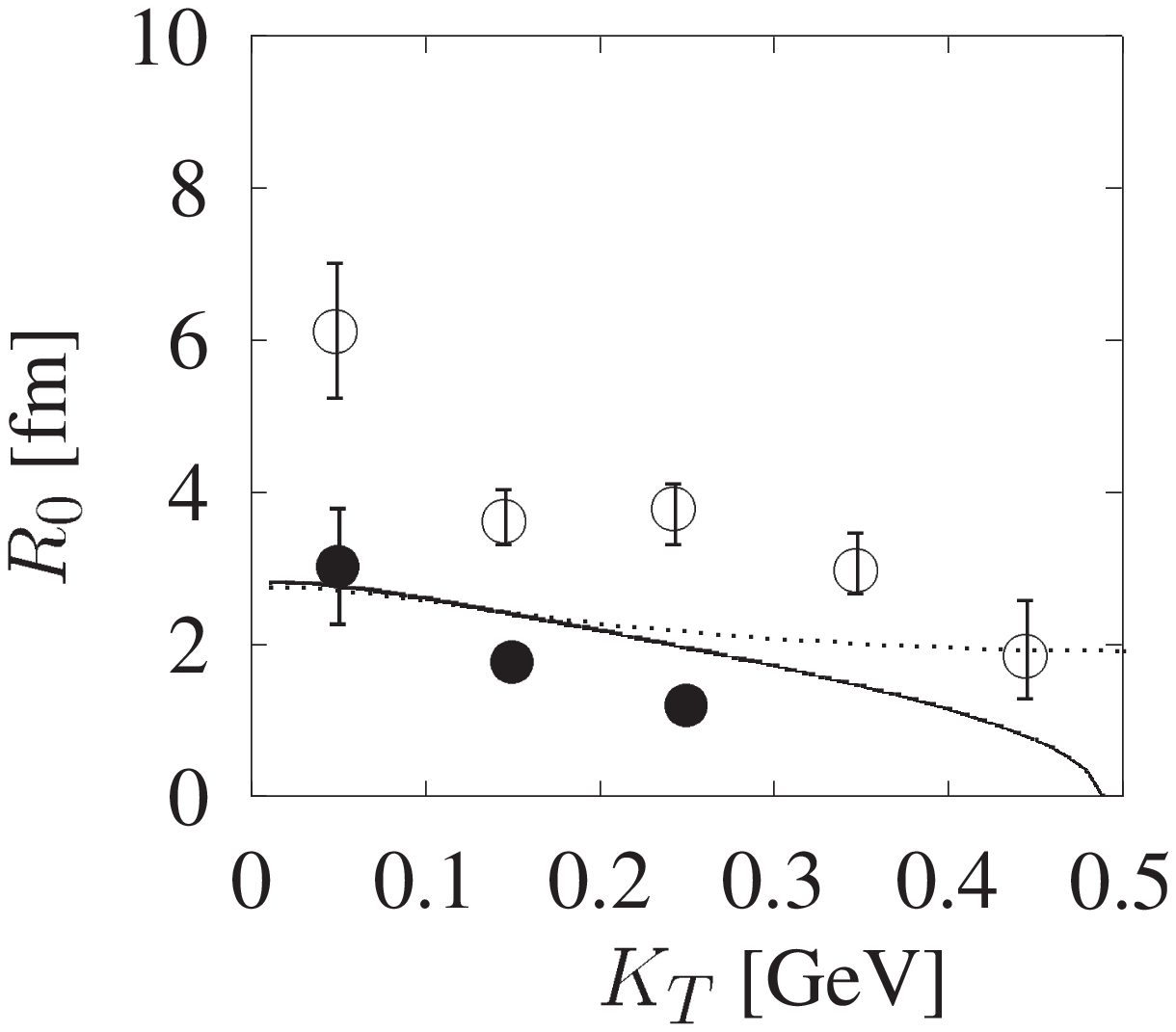}\\
  \end{center}
  {\small Fig.~1 $K_T$ dependence of YKP radii. Open and closed circles show
  the experimental data and HBT radii, respectively. Solid lines stand for
  space-time extensions  (\protect\ref{rt})-(\protect\ref{rl}). Dotted lines
  stand for source sizes. ($\Delta z$ for $R_\|$ and  $\Delta t$ for
  $R_0$.)}
 \end{minipage}\\

Figure 1 shows $K_T$ dependence of the pion HBT radii. Close circles stand
for the HBT radii. Solid lines are space-time extensions which are expected
to agree with the HBT radii when only thermal pions are considered. Dotted
lines stand for the source sizes ($\Delta z$ for $R_\|$ and $\Delta t$
for $R_0$). As far as $R_\bot$ and $R_\|$ are concerned, those quantities
seem to be consistent with one another and well reproduce the NA49
experimental results \cite{na49} (open circles). On the other hand, $R_0$ is
smaller than the experimental one in spite of the fact our calculation
includes a first order phase transition. The time durations $\Delta t$ agrees
with $R_0$ at small $K_T$. However, some deviations from $R_0$ can be seen at
large $K_T$ region, that means a failure of interpretation of $R_0$ as the
time duration. We find that this difference occurs from the third term in
Eq.(\ref{r0}). 

In Eq.(\ref{r0}), the two additional terms means correlation between $x$ and
$z$ (the second term) and the thickness of the source (the third
term). Though contribution from the second term is small in our model, the
third term becomes negative in same order with $(\Delta t)^2$
\cite{paper}. Such a kind of the source is called an ``opaque source''. The
``opaque'' source \cite{opaque} emits particles dominantly from the surface
and the ``transparent'' source emits from whole region. Figure 2 shows the
source function (\ref{S}) projected onto the $x-y$ plane,
$\tilde{S}_T(x,y)=*\int dz dt S(x^\mu,K^\mu)$ being normalized as $\int dx dy
\tilde{S}_T(x,y)=1$. Note that the average transverse momentum of the emitted
pions is parallel to $x$ direction in the figure. The left figure shows
\textit{clearly} surface emission; pions are emitted from the crescent region
and the source is thinner in the $x$ direction than in the $y$
direction. This is the typical property of the opaque source which appears in
the factor $(\Delta x)^2-(\Delta y)^2$ in Eq.(\ref{r0}). Note that this
factor also appears in the longitudinal components of the size,
Eq.(\ref{rl}), but the prefactor $\beta_\|^2$ makes the contribution to
$R_\|^2$ small. If we neglected the transverse flow by artificially putting
$U^r=0$, the source function becomes as shown in the right figure in
Fig.~2. In this case, the source function is proportional to the space-time
volume of freeze-out hypersurface. As a consequence of restoration of the
azimuthal symmetry which is shown clearly in the figure, the above measure of
source opacity vanishes. Although the particle emission takes place almost
from the thin surface, the source is not opaque in this sense. When the
transverse flow exists, the source function is deformed by thermal Boltzmann
factor $\exp(K_T \sinh Y_T \cos \phi /T_f)$ with $Y_T$ almost proportional to
$r$. Consequently, the number of emitted particles increases in the region
where $x>0$ (i.e., $\cos \phi >0$) and decreases in the region where $x<0$
($\cos \phi <0$). This flow effect deforms the surface dominant volume
distribution (right figure) to the crescent shape (left figure). 

\vspace*{1cm}
  \begin{minipage}[c]{13cm}
   \baselineskip=12pt
   \begin{center}
	\includegraphics[scale=0.6]{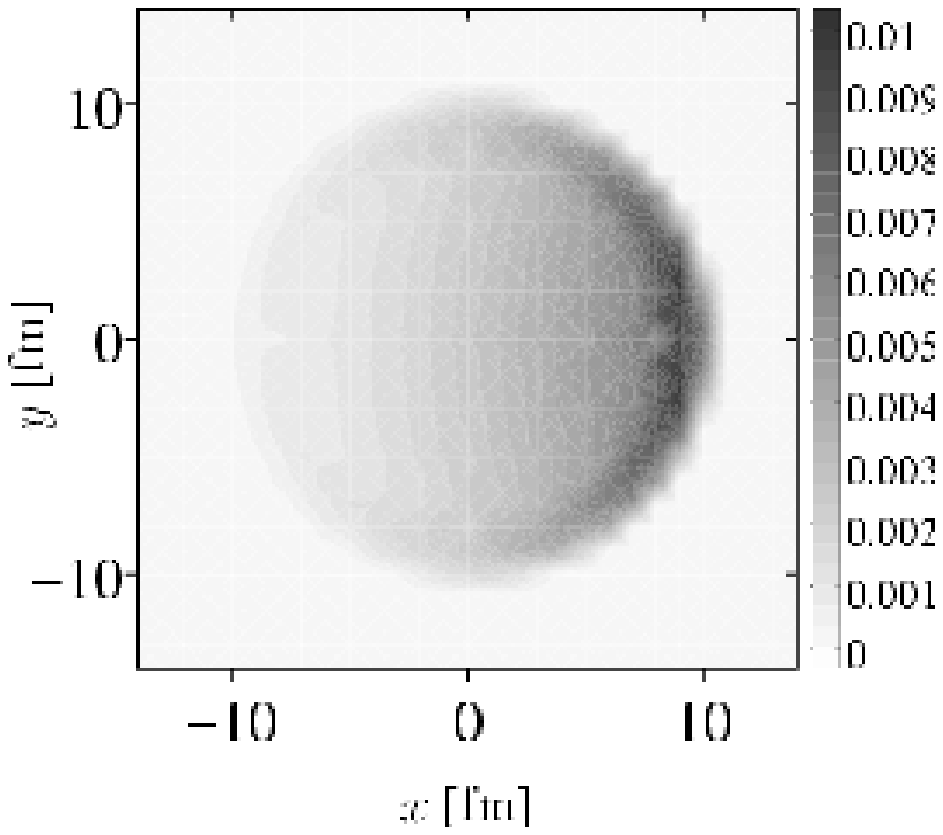}\,
	\includegraphics[scale=0.6]{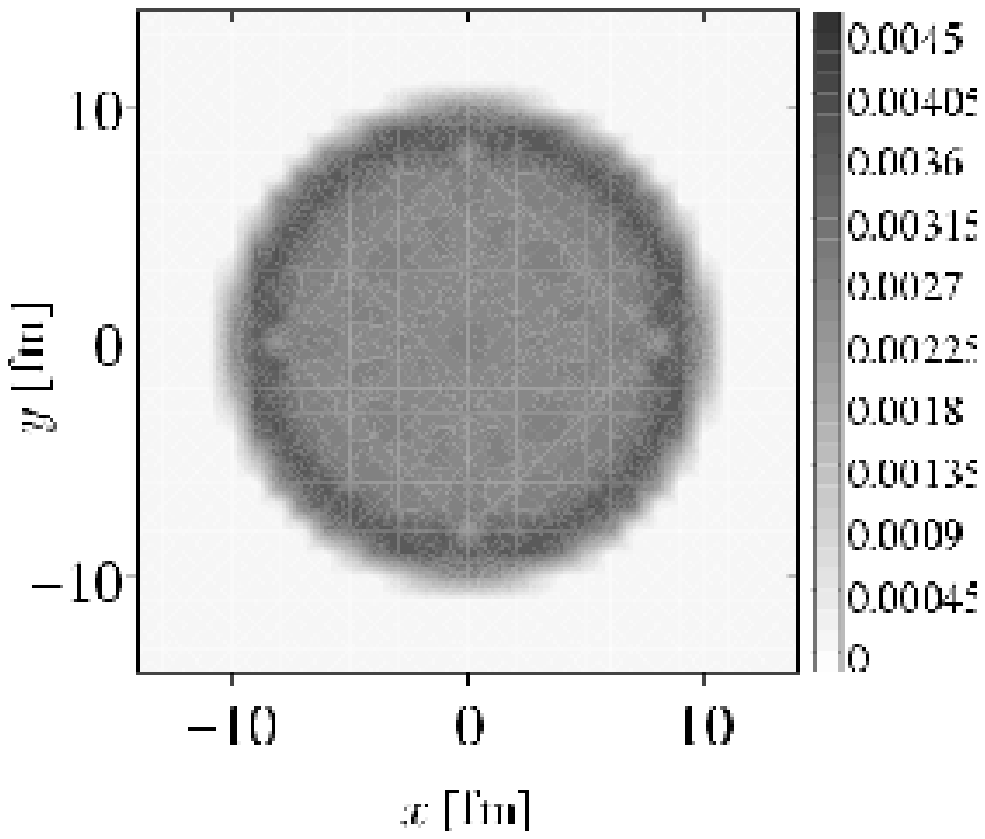}
   \end{center}
  {\small Fig.~2 The source functions as functions of transverse coordinates
   $x$ and $y$ for $K_T=450$ MeV and $Y_{\pi\pi}=4.15$
   ($Y_{\pi\pi}=\frac{1}{2}\ln \frac{K^0+K_L}{K^0-K_L}$). The left figure
   denotes the source function with the transverse flow. The right figure
   denotes the source function \textit{without} transverse flow.}
  \end{minipage}\\

In summary, we analyze $K_T$ dependence of source parameters of the YKP
parametrization based on the relativistic hydrodynamics for the CERN-SPS 158
GeV/A Pb+Pb collisions. We obtain the results which are almost consistent
with the experiment. However, the temporal source parameter, $R_0$, shows
smaller value than the experiment. The source opacity makes the
interpretation of $R_0$ as the time duration doubtful at large $K_T$ region. We
find that the source opacity is caused by the transverse flow and the
characteristics of the freeze-out hypersurface, the surface dominant
freeze-out. The deviations of our results from the experiment would be
improved by including the resonance decay and other effects.

The authors are indebted to Professor I.~Ohba and Professor H.~Nakazato for
their helpful comments. This work was partially supported by a Grant-in-Aid
for Science Research, Ministry of Education, Science and Culture, Japan
(Grant No. 09740221) and Waseda University Media Network Center.

\end{document}